\documentclass[twocolumn,prl,aps,showpacs]{revtex4}

\usepackage{subfigure}
\usepackage{psfrag,graphicx}
\usepackage{dcolumn}
\usepackage{amsmath,amssymb}
\usepackage{bm}
\usepackage[dvips]{color}
\usepackage{overpic}

\definecolor{mygrey}{gray}{0.35}
\definecolor{mygreen}{rgb}{0.85,1,0.9}
\definecolor{myzard}{cmyk}{0,0,0.05,0}
\definecolor{mywhite}{rgb}{1,1,1}
\definecolor{myred}{rgb}{1,0,0}

 \def\ee{\mathord{\rm e}}
 
 \def\ii{\mathord{\rm i}}

\def\half{\textstyle\frac{1}{2}}

\renewcommand{\ii}{{\rm i}}
\renewcommand{\ee}{{\rm e}}

 \newcommand{\ket}[1]{|#1\rangle}

\begin{document}

\title[Short Title]{Topology induced anomalous defect production by crossing a quantum critical point}

\author{A. Bermudez$^1$, D. Patan\`e$^{2,3}$, L. Amico$^{2,3}$, and M. A. Martin-Delgado$^{1}$}

\affiliation{$^1$Departamento de F\'{i}sica Te\'orica I,
Universidad Complutense, 28040 Madrid, Spain \\
$^2$MATIS-INFM  $\&$ Dipartimento di Metodologie Fisiche Chimiche(DMFCI), 
Universit\`{a} di Catania, viale A. Doria 6, 95125 Catania, Italy\\
$^3$Departamento de F\'{i}sica de Materiales, Universidad Complutense,
28040 Madrid, Spain}

\pacs{64.60Ht, 73.20.At, 73.43.Nq, 11.15.Ha }
\begin{abstract}
We study the influence of topology on the quench dynamics of a system driven across a quantum critical point. 
We show how the  appearance of certain edge states, which fully characterise  the topology of the system, dramatically modifies the 
process of defect production during the crossing of the critical point.  Interestingly enough, the density of defects  is no longer described by the Kibble-Zurek scaling, but determined instead by the non-universal topological features of the system. Edge states are shown to be robust against defect production, which highlights their topological nature.

\end{abstract}

\maketitle

Phase transitions occur between two different states of matter with  different orders. When a critical point is crossed with a certain rate, the order can be established only partially, and  the  system displays a non-vanishing density of defects.  The  Kibble-Zurek (KZ) mechanism is the paradigm to analyze the  dynamics of a system driven across a phase transition$\text{\cite{Kibble,Zurek}}$. Accordingly, the density of defects produced during the crossing is uniquely determined by the  universality class of the system. Remarkably enough, this  mechanism accurately describes also quantum phase transitions, occurring at absolute zero temperature${\text{\cite{KZ1,KZ2, KZ3, KZ4, KZ5, KZ6, KZ7, KZ8, KZ9}}}$; the effect of crossing a quantum critical line was shown in ${\text{\cite{KZline,sengupta}}}$. Here we show that the topology of the system  may induce an anomalous defect production which strongly deviates from the KZ scaling law. We demonstrate that  such deviation is caused by the properties of the characteristic edge states, which are a fingerprint of the system non-trivial topology.

The effect of boundary conditions  is peculiar
 for the class of systems possessing a topological order \cite{wen_book}. 
In fact, such systems are characterized by a ground state degeneracy 
that strongly depends on the topology of the of the manifold on which 
they live. 
Besides, a hallmark  of topological order  is the appearance of states 
localised at the boundary of the system, the so-called edge states.  
Edge states emerge naturally in a wide variety of systems, 
such as the fractional quantum Hall effect~\cite{wen_fqh}, 
one-dimensional spin models~\cite{aklt}, or  Majorana fermions on a lattice~\cite{kitaev}. 
Recently, edge states have been realized experimentally in 
topological insulators connected with quantum spin Hall effect~\cite{koenig, hsieh}.

The subject of this letter is to investigate the effects of topology 
in the out-of-equilibrium dynamics induced by the crossing of a quantum critical point. 
In this context, the preparation of topological order via adiabatic evolution was addressed in \cite{hamma}. 
Moreover, the scaling of defects created by crossing a line of quantum critical points of topological nature 
was studied in \cite{sengupta}.
The scenario that emerged supports the KZ mechanism, with a generalized scaling law that takes into account the infinite number of critical points crossed. Let us remark that no peculiar behavior induced by the topological nature of the system has been found so far. 
In the following we will show how  edge states dramatically modify the KZ scaling law, thus indicating the need of  a new paradigm to describe the dynamics of topological defect production. 
To address this problem we consider a specific model originally proposed  by Creutz in \cite{creutz}.
As we shall discuss below, this system offers a rich scenario where to study the influence of  topology on the defect production across a quantum critical point, and may serve as a paradigmatic model for more general topological systems.

\begin{figure}[!hbp]

\centering

\begin{overpic}[width=5.0cm]{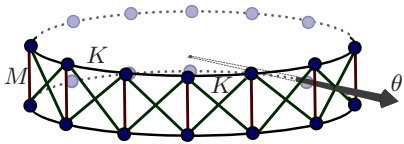}
\put(17,30.5){$K$}
\put(-5,25){$M$}
\put(50,22.5){$K$}
\put(98,23){$\theta$}
\end{overpic}
\caption{Diagram of the Creutz ladder with periodic boundary conditions: Electrons freely move across the ladder hoping along horizontal, vertical and diagonal links that join the different sites of the lattice.}\label{creutz_ladder_scheme}

\end{figure}

The Creutz model describes the dynamics of a spinless electron moving in the ladder system depicted in Fig.~\ref{creutz_ladder_scheme}, as dictated by the following Hamiltonian
\begin{equation}
\label{creutz_hamiltonian_b_field}
\begin{split}
H:=&-\sum_{n=1}^{L}\left[K\left(\ee^{-\ii\theta}a^{\dagger}_{n+1}a_n+\ee^{\ii\theta}b^{\dagger}_{n+1}b_n\right)+\right . \\
&\left. +K\left(b^{\dagger}_{n+1}a_n+a^{\dagger}_{n+1}b_n\right)+M a^{\dagger}_{n}b_n+\text{h.c.}\right],
\end{split}
\end{equation}
where we have introduced the fermionic operators $a_n$, and $b_n$ associated to the $n-$th site of the upper and lower  chain respectively. The hopping along  horizontal and diagonal links  is described by the parameter $K$, and the vertical one by $M$; additionally  a magnetic flux $\theta\in[-\pi/2,\pi/2]$ is induced by a perpendicular magnetic field. 

For small vertical hopping $M<2 K$,  the model has a second order quantum phase transition at   $\theta_c=0$ with   equilibrium critical exponents  $\nu=z=1$ (see Eq. \ref{periodic_spectrum}). The quantum critical point separates two band insulators with topologically distinct configuration of currents (leading to opposite circulation of thereof along the ladder). We thus focus on a quench of the magnetic flux $\theta(t):=v_qt-\pi/2\in[-\pi/2,\pi/2]$ at constant rate $v_q$.
According to the KZ mechanism, as the system crosses the critical point and the energy gap vanishes, excitations are unavoidably produced as $P^{\text{KZ}}_{\text{def}}\propto v_q^{\frac{d\nu}{z\nu+1}}$, 
where $d$ is the system dimension. We now explore the influence of the ladder topology on the dynamics across a quantum critical point and describe the circumstances where strong deviations from the KZ scaling arise.

\begin{figure}[!hbp]

\centering
\subfigure[\hspace{1ex}Energy spectrum for the periodic ladder]{
\begin{overpic}[width=6.1cm]{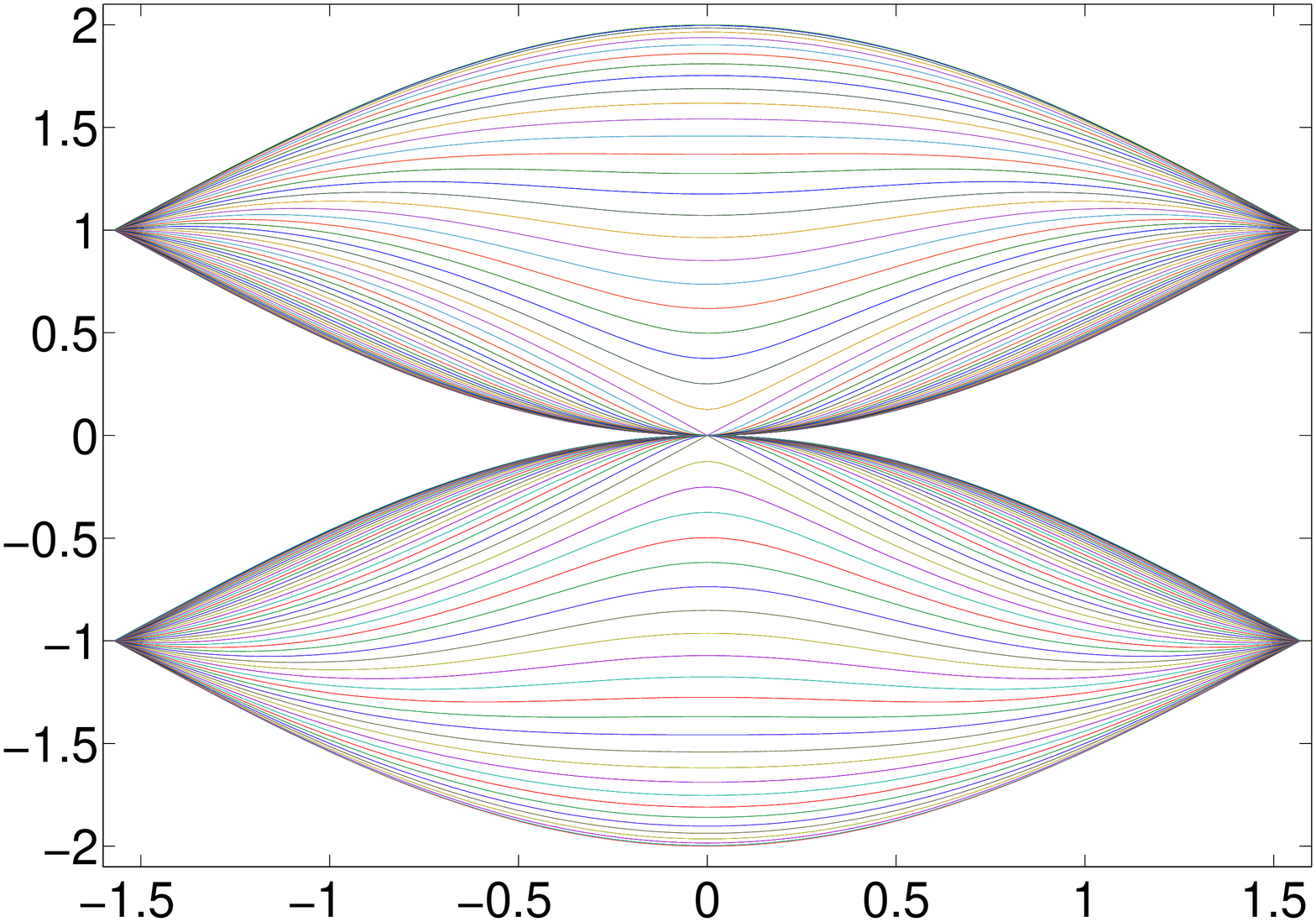} \label{theta_periodic_energy_spectrum}
\put(-5,60){$E/2K$}
\put(93,3){$\theta$}
\end{overpic}
}
\subfigure[\hspace{1ex}Energy spectrum for the open ladder]{
\begin{overpic}[width=6.1cm]{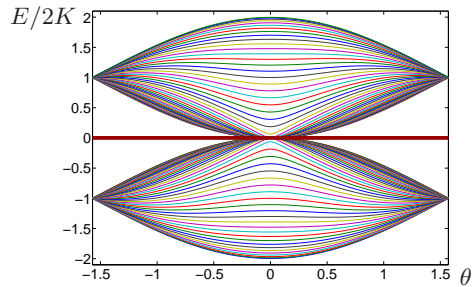} \label{theta_open_energy_spectrum}
\put(-5,60){$E/2K$}
\put(93,3){$\theta$}
\end{overpic}
}
\caption{ Energy spectrum of the  Creutz ladder for $m=0$  as a function of the magnetic flux $\theta$: (a) For periodic boundary conditions we observe that the energy gap vanishes at $\theta_c=0$, which correspond to a quantum critical point. (b) For open boundary conditions we have two additional zero-energy levels which correspond to edge states that characterize the topology of the ladder.}
\end{figure}

Let us first discuss the quench dynamics of a ladder with periodic boundary conditions $a_{L+1}=a_1$, $b_{L+1}=b_1$, which impose a closed topology on the system. Introducing the fermionic operators in momentum space   $a_q:=\sum_n a_n\ee^{\ii q n}$, and $ b_q:=\sum_n b_n\ee^{\ii q n}$, where $q:=\frac{2\pi}{L}j$  with $j=1,...,L$, the  Hamiltonian in equation~\eqref{creutz_hamiltonian_b_field} reads
\begin{equation}
\label{creutz_hamiltonian_b_field_momentum}
H=-2K\sum_q(a^{\dagger}_q,b^{\dagger}_q)\left(\begin{array}{cc}  \cos(q-\theta) & m+\cos q \\ m+\cos q &  \cos(q+\theta)  \end{array}\right)\left(\begin{array}{c} a_q \\b_q  \end{array}\right),
\end{equation}
where  $m:=M/2K$ is the relative vertical hopping, and $K$ fixes the energy scale. Hence, the phase diagram depends upon the parameters $m$ and $\theta$ with the following energy spectrum $E_{q\pm}:=E/2K$ 
\begin{equation}
\label{periodic_spectrum}
E_{q\pm}=-\cos q\cos \theta\pm\sqrt{\text{sin}^2q\sin^2\theta+(m+\cos q)^2}.
\end{equation}
In the following  we will focus on  the regime with $m=0$  (see fig.~\ref{theta_periodic_energy_spectrum}). Let us notice however that  the results obtained are robust under small fluctuations of $m<1$.

\begin{figure}[top]
\centering
\subfigure[\hspace{1ex}Defect production in the periodic ladder]{
\begin{overpic}[width=7.0 cm]{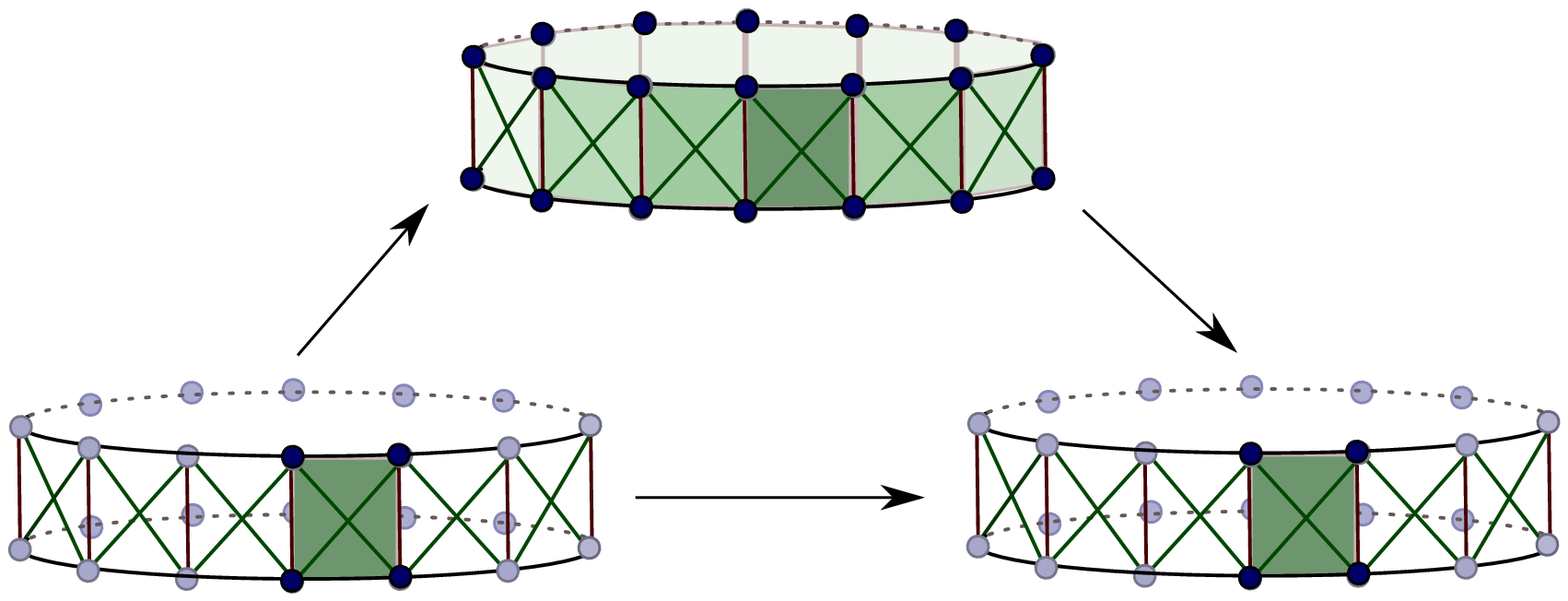} \label{plaquette_theta_quench}
\end{overpic}
}
\subfigure[ \hspace{1ex}Defect production in the open ladder]{
\begin{overpic}[width=7.0 cm]{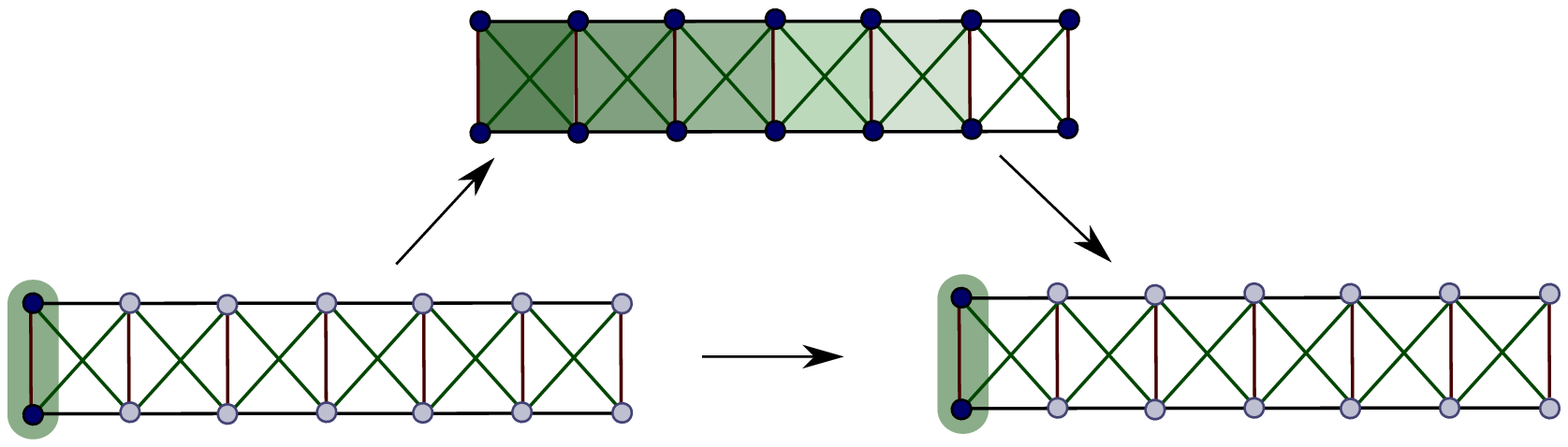} \label{edge_state_m_quench}
\end{overpic}
}
\caption{Scheme of the production of delocalised defects in the magnetic flux  quench $\theta(t)$: (a) In the periodic case, starting from the localised plaquette $\ket{-2K}_{n}$, delocalised defects are produced as the system is driven across the critical point $\theta_c=0$. At the end of the quench $\theta\to\pi/2$, the ladder will be in a linear superposition of the plaquette-like ground state and de-localised excited states. (b) For open boundary conditions, we have to consider the dynamical quench of the topological edge state localised at the left boundary of the ladder. }
\end{figure}

We shall consider an initial ground state  corresponding to $\theta(0)=-\pi/2$ and localised within a plaquette at $(n,n+1)$,
$\ket{-2K}_{n}:=\half(-\ii a^{\dagger}_n+b^{\dagger}_n+a^{\dagger}_{n+1}-\ii b^{\dagger}_{n+1})\ket{0} $. Such  state arises due to the destructive interference between different hopping paths, which forbids electron tunneling to next-nearest neighboring rungs $n\nrightarrow n\pm 2$. 
At the end of the quench $\theta(t_{\text{f}})=+\pi/2$, 
the system shall be found in a state $|\Psi(t_{\text{f}})\rangle$ given by the superposition of 
the  ground state still localised within the plaquette (adiabatic evolution of the initial state), and de-localized defects which have been excited close to the critical point. This dynamics, schematically depicted in fig.~\ref{plaquette_theta_quench},  can be rigorously described as a collection of $L$  Landau-Zener processes~\cite{LZ,LZ_bis} (see eq.~\eqref{creutz_hamiltonian_b_field_momentum}). The total density of excitations at the end of the quench is $P_{\text{def}}=\sum_{E>0}|\langle E | \Psi(t_{\text{f}})\rangle|^2$, $\langle E|$ being the eigenvectors of the Hamiltonian. 

\begin{figure}[top]

\subfigure[\hspace{1ex} PBC: Dynamical quench of a bulk plaquette]{
\begin{overpic}[width=7.0 cm]{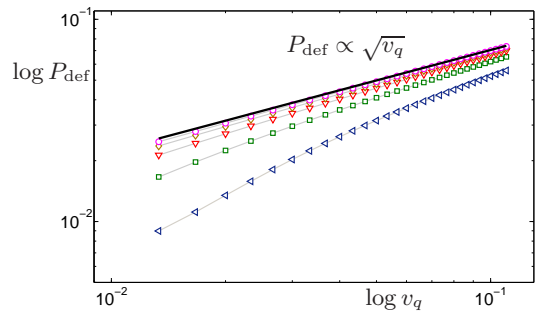} \label{kz_quench}
\put(65,2){$\log v_q$}
\put(-2,45){$\log P_{\text{def}}$}
\put(50,50){$ P_{\text{def}}\propto \sqrt{v_q}$}
\end{overpic}
}

\centering
\subfigure[\hspace{1ex} OBC: Dynamical quench of a bulk plaquette]{
\begin{overpic}[width=7.0cm]{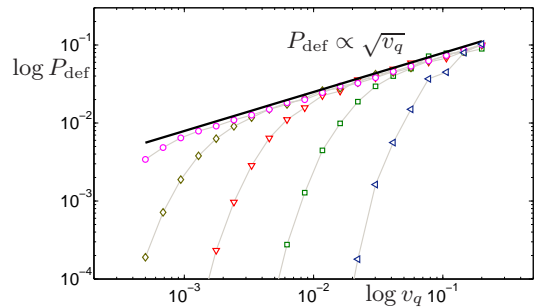} \label{theta_kz_bulk_plaq}
\put(65,2){$\log v_q$}
\put(-2,45){$\log P_{\text{def}}$}
\put(50,50){$ P_{\text{def}}\propto \sqrt{v_q}$}
\end{overpic}
}
\centering
\subfigure[\hspace{1ex} OBC: Dynamical quench of an edge plaquette]{
\begin{overpic}[width=7.5cm]{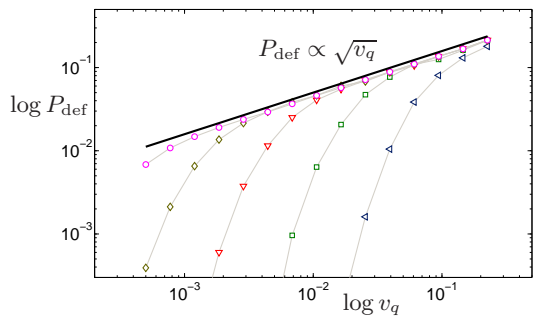} \label{theta_kz_edge_plaq}
\put(60,2){$\log v_q$}
\put(1,37){$\log P_{\text{def}}$}
\put(45,47){$ P_{\text{def}}\propto \sqrt{v_q}$}
\end{overpic}
}

\subfigure[\hspace{1ex}OBC: Dynamical quench of an edge state]{
\begin{overpic}[width=7.0 cm]{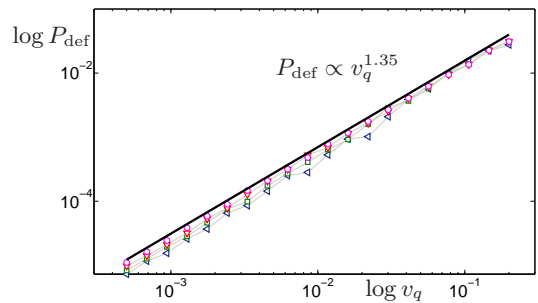} \label{theta_kz_edge_state}
\put(65,2){$\log v_q$}
\put(-2,50){$\log P_{\text{def}}$}
\put(48,44){$ P_{\text{def}}\propto v_q^{1.35}$}
\end{overpic}
}
\caption{ Scaling of the density of defects produced during a magnetic flux quench $\theta(t)$.  Periodic boundary conditions: (a) density of defects $P_{\text{def}}$  produced after the quench of the magnetic flux as a function of the quench rate $v_q$ in logarithmic scale for a ladder with $L\in\{80,160,320,640,1280\}$ rungs for an initial plaquette-like state. In the thermodynamic limit, the density of defects fulfills the KZ scaling  $P_{\text{def}}\propto\sqrt{v_q}$. Open boundary conditions: (b)  plaquette localised  in the bulk $\ket{-2K}_{L/2}$ for  $L\in\{10,20,40,80,160\}$ rungs,  (c) plaquette localised  in the  boundaries $\ket{-2K}_{1}$ for  $L\in\{10,20,40,80,160\}$ rungs, (d) edge state $\ket{l}$  for  $L\in\{10,20,40,80,160\}$ rungs . }
\end{figure}

In fig.~\ref{kz_quench}, we present the density of produced defects as the thermodynamic limit is approached. It readily follows  that the density of defects is accurately described by the  scaling $P_{\text{def}}\propto\sqrt{v_q}$, which is in complete agreement with the KZ prediction for the critical exponents of the model. 
In fact, despite the non linear nature of the  quench of $\theta$, equation\eqref{creutz_hamiltonian_b_field_momentum} can be linearized close to gapless mode $q_g=\pi/2$ at $\theta_c$, yielding thus the conventional scenario where the KZ scaling holds.   
 
Once the periodic ladder has been thoroughly described, we may study the influence of a different topology on the quench dynamics of the ladder. In the case of open boundary conditions $a_{L+1}=0$, $b_{L+1}=0$, the translational invariance of the system is lost. The dynamics can no longer be described as a collection of $L$ uncoupled two-level systems associated to a given mode $q$, and transitions between all energy levels occur.  
We observe that the quantum phase transitions  in  fig.~\ref{theta_open_energy_spectrum}  still resembles the periodic case in fig.~\ref{theta_periodic_energy_spectrum}. However, the opening of the chain is crucially accompanied by the appearance of two new levels whose energies lie close to zero. Such levels, which correspond to edge states pinned at the ends of the lattice,  are a direct consequence of the ladder topology (i.e. whether it is an open quasi-one-dimensional chain, or a closed quasi-one-dimensional ring). Below, we study the effects of these new states, and thus the effects of topology, in the ladder quench dynamics.

For open boundary conditions, we can initialise the system in two qualitatively different localised states, namely, the plaquette-like state  localised at the bulk (or boundary), or the  topological  edge state located at either side of the ladder $\ket{l}:=\textstyle{\frac{1}{\sqrt{2}}}(a_1^{\dagger}-\ii b_1^{\dagger})\ket{0}$, and $
\ket{r}:=\textstyle{\frac{1}{\sqrt{2}}}(-\ii a_L^{\dagger}+b_L^{\dagger})\ket{0}$ (see fig.~\ref{edge_state_m_quench}).
In the case in which the system is initialised in a plaquette-like state,
the density  of defects scales according to the KZ mechanism $P_{\text{def}}\propto\sqrt{v_q}$.
Being plaquette-like states independent on the ladder topology,
we consistently  find the same scaling obtained  in the periodic case.
Moreover, KZ scaling holds regardless of the plaquette 
position within the ladder, either located at the bulk (fig.~\ref{theta_kz_bulk_plaq}) or edge (fig.~\ref{theta_kz_edge_plaq}). 
These results rule out the existence of a non-topological attraction mechanism 
at the boundaries, which would modify the ladder 
quench dynamics.

Let us now focus on the quench dynamics of a topological edge state localised in the left side of the ladder $\ket{l}$. Interestingly enough, the energy gap between this edge state and the remaining excitations shows a similar scaling as the gap of a localised plaquette, and thus the KZ mechanism would results in a similar scaling $P^{\text{KZ}}_{\text{def}}\propto\sqrt{v_q}$. Contrarily, we have found
(see Fig.~\ref{theta_kz_edge_state}) the surprising result that the scaling of the density of defects for an initial topological edge state deviates pronouncedly from the KZ theory 
\begin{equation}
\label{non_KZ_scaling}
P_{\text{def}}\propto v_q^{1.35}.
 \end{equation}
Accordingly, edge states are more robust (with respect to non-topological states) because a smaller amount of defects is generated during adiabatic quenches $v_q\ll1$.  This anomalous scaling is due to topological constraints that decouple the edge states from the excitations that lie close to the gapless mode, whose universal density of states gives rise to the KZ scaling. Therefore, we can state that the quench dynamics of these topological edge states is not dictated by the universality class of the system, but rather induced by  non-universal effects of the  topology. It is precisely this fact which does not allow a theoretical prediction of  Eq.~\eqref{non_KZ_scaling} based on the critical properties of the quantum phase transition.

Summarizing, in this letter we have presented a detailed study of the effect of the topology on the  quench dynamics. We have shown that the influence of edge states, which constitute a clear signature of the system topology, leads to a peculiar scaling of quench defects. In this respect, we have presented an insightful model, the so-called Creutz ladder, where the influence of these edge states on the quench dynamics can be neatly described. As we noticed, an adiabatic quench of  the magnetic flux produces a smaller amount of defects when the ladder is initially in a topological edge state. 
Being the  presence of in-gap edge states  a generic feature  of  topological band insulators \cite{kane,fu,lee}, we expect that an anomaly in the defect production could emerge  beyond the specific model we considered.
We would like to conclude the letter commenting on the relevance of our results for quantum information. The suppression of the  defects production law suggests that the  edge states are particularly  suitable candidates to implement a protected quantum memory, even in the presence of strong fluctuation of a magnetic field that may drive the system across a critical point. Let us finally note that a natural candidate to experimentally implement this topological quench dynamics is that of spinor Fermi gases in optical lattices~\cite{ol_review}. The two species of fermionic operators in the ladder may correspond to the fermionic operators associated to different spin-components, whereas the non-interacting regime is reached by means of Feschbach resonances. 

{\it Acknowledgments.} A.B. and M.A.M.D  acknowledge financial support 
from the Spanish MEC project FIS2006-04885, the 
project CAM-UCM/910758, and, together with D.P, the ESF Science Programme INSTANS 2005-2010. Additionally, A. B.
acknowledges support from a FPU MEC grant.


\end{document}